\documentclass[prl,twocolumn,showpacs,preprintnumbers,amsmath,amssymb, superscriptaddress]{revtex4}
\usepackage{graphicx}
\usepackage{dcolumn}
\usepackage{bm}
\usepackage{fancyhdr}
\usepackage{float}
\usepackage{ifthen}
\usepackage{eurosym}

\begin{document}
\preprint{Yb3Rh4Sn13}
\title{Distinct vortex-glass phases in Yb$_{3}$Rh$_{4}$Sn$_{13}$ at high and low magnetic fields}

\ \author{D.~G.~Mazzone}
\ \affiliation{Laboratory for Neutron Scattering and Imaging, Paul Scherrer Institut, 5232 Villigen PSI, Switzerland}
\
\ \author{J.~L.~Gavilano}
\ \affiliation{Laboratory for Neutron Scattering and Imaging, Paul Scherrer Institut, 5232 Villigen PSI, Switzerland}

\ \author{R.~Sibille}
\ \affiliation{Laboratory for Developments and Methods, Paul Scherrer Institut, 5232 Villigen PSI, Switzerland}

\ \author{M.~Ramakrishnan}
\ \affiliation{Laboratory for Neutron Scattering and Imaging, Paul Scherrer Institut, 5232 Villigen PSI, Switzerland}
\ \affiliation{Swiss Light Source, Paul Scherrer Institut, 5232 Villigen PSI, Switzerland}

\ \author{C.~D.~Dewhurst}
\ \affiliation{Institut Laue-Langevin, 71 Avenue des Martyrs, 38000 Grenoble, France}

\ \author{M.~Kenzelmann}
\ \affiliation{Laboratory for Developments and Methods, Paul Scherrer Institut, 5232 Villigen PSI, Switzerland}


\date{\today}

\begin{abstract}
The vortex lattice (VL) in the mixed state of the stannide superconductor Yb$_{3}$Rh$_{4}$Sn$_{13}$ has been studied using small-angle neutron scattering (SANS). The field dependencies of the normalized longitudinal and transverse correlation lengths of the VL, $\xi_L/a_0$ and $\xi_T /a_0$, reveal two distinct anomalies that are associated with vortex-glass phases below  $\mu_0H_l$~$\approx$~700~G and above $\mu_{0}H_h$~$\sim$~1.7~T ($a_0$ is the intervortex distance). At high fields, around 1.7~T, the longitudinal correlation decreases abruptly with increasing fields indicating a weakening (but not a complete destruction) of the VL due to a phase transition into a glassy phase,  below $\mu_{0}H_{c_2}$(1.8 K)~$\approx$~2.5~T. $\xi_L/a_0$ and $\xi_T /a_0$, gradually decrease for decreasing fields of strengths less than 1~T and tend towards zero. The shear elastic modulus $c_{66}$ and the tilting elastic modulus $c_{44}$ vanish at a critical field $\mu_0H_l$~$\approx$~700~G, providing evidence for a disorder-induced transition into a vortex-glass.  A 'ring' of scattered intensity is observed for fields lower than 700~G, $i.e.$, $\mu_{0}H_{c_1}$~=~135~G~$<$~$\mu_{0}H$~$<$~700~G. This low-field phenomenon is of different nature than the one observed at high fields, where  $\xi_L/a_0$ but not $\xi_T/a_0$, decreases abruptly to an intermediate value.
\end{abstract}


\maketitle

\bigskip

Vortex matter in type-II superconductors has been the subject of many investigations, in particular since the discovery of high-$T_{c}$ Cuprates \cite{Blatter, LeDoussal}. There, it became evident that quantum fluctuations and dimensionality play an important role in many new and interesting physics involving the vortex lattice. Novel vortex matter phenomena have been observed, such as glassy phases, dimensional crossover and vortex lattice melting due to thermal and quantum fluctuations \cite{Brandt, Lee}. In spite of the large research effort, many open questions still remain even in simple conventional superconductors. This is partly due to the very fine sensitivity of the vortex lattice to details of the Fermi surface, and (microscopic) pinning of vortices to the underlying crystal lattice. The effects of pinning are difficult to calculate, but it takes a dominant role under some circumstances. An experimental technique very adequate to study the vortex lattice in superconductors is small-angle neutron scattering (SANS), which is a direct probe of the reciprocal lattice of the vortices. We used this technique to study the VL in Yb$_{3}$Rh$_{4}$Sn$_{13}$ and found some unusual results described below.

Yb$_{3}$Rh$_{4}$Sn$_{13}$ crystallizes in a cubic lattice with tetragonal point group symmetry consisting of a threefold rotation axis along the cube diagonals but without fourfold rotation axis (space group $Pm$-3$n$) with a unit cell containing  40 atoms and two formula units, $Z$~=~2 \cite{Remeika}. Recent observations of interesting physical phenomena for some members of this family resulted in a surge of interest of these materials \cite{Klintberg, Gerber, Levett}. The Ytterbium atom in Yb$_{3}$Rh$_{4}$Sn$_{13}$ is in a mixed valence state \cite{miraglia}. However, the Sommerfeld coefficient $\gamma$ of this material displays only an enhancement of a factor of four compared to the case of the isostructural compound Ca$_{3}$Ir$_{4}$Sn$_{13}$ \cite{Wang, Sato}. It has been claimed that a broad transition into a vortex glass involving multiple steps occurs below $H_{c_2}$ \cite{Sato, Tomy}.

\begin{figure}[tbh]
\includegraphics[width=\linewidth]{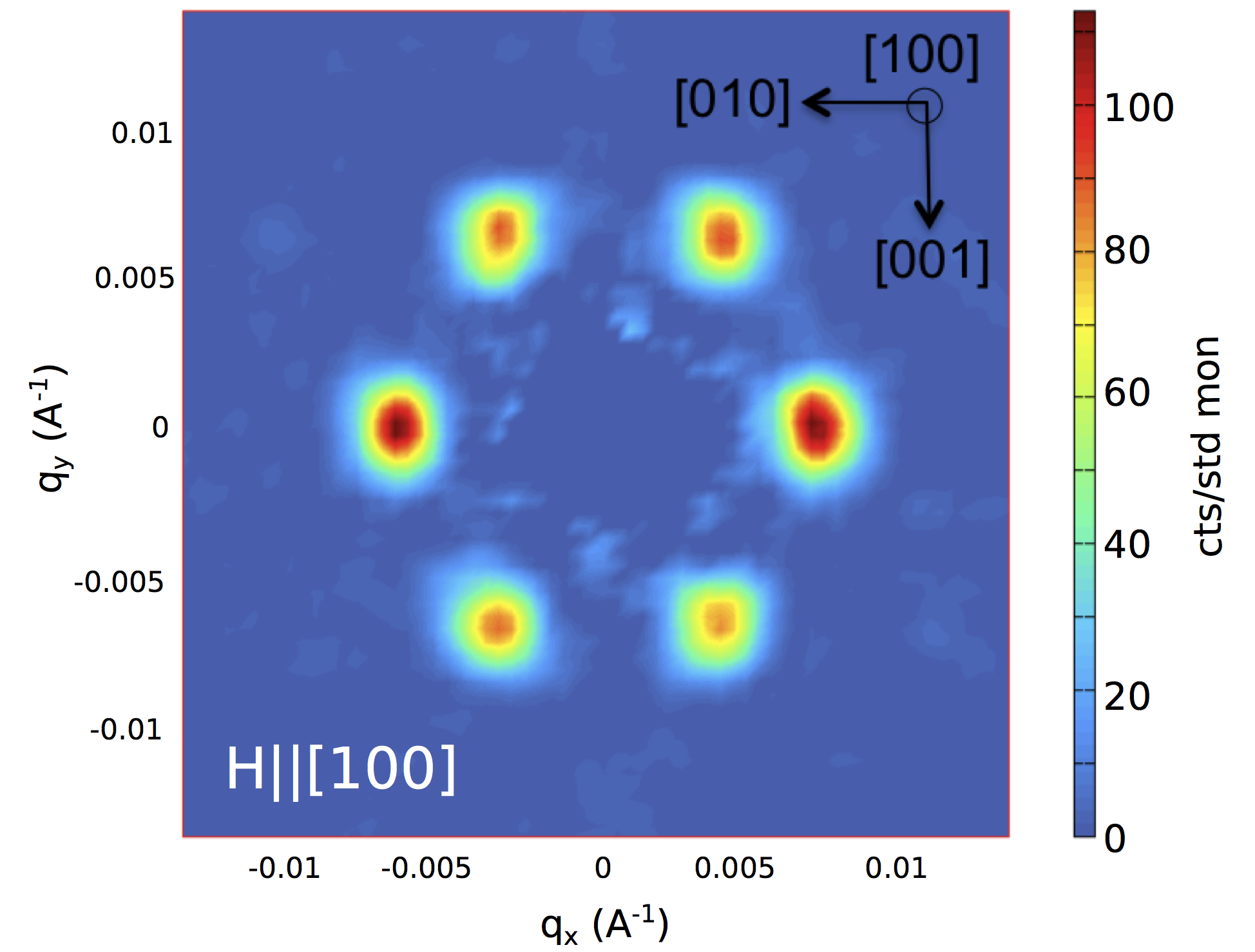}
\caption{Vortex lattice diffraction pattern at $T$~=~50~mK and $\mu_{0}H$~=~3500~G for $\vec{H}||[100]$.}
\label{phasediagramm}
\end{figure}

Here, we present results of SANS measurements in the mixed state of Yb$_{3}$Rh$_{4}$Sn$_{13}$ at 50~mK and 1.5~K. The magnetic field strengths were between 350 and 18500~G and applied along the crystallographic [100] axis. Our SANS data reveal a single-domain VL structure in the mixed state of Yb$_{3}$Rh$_{4}$Sn$_{13}$ for fields below 1.85~T (see Fig. \ref{phasediagramm}). For low fields (above $H_{c_1}$) and high fields (below $H_{c_2}$) we observe rapid changes of the normalized correlation lengths of the vortices, longitudinal $\xi_{L}/a_{0}$ and transverse $\xi_{T}/a_{0}$. $a_{0}$, the distance between the vortices, decreases with increasing field. These changes are interpreted as evidence for different phase transitions. At high fields, 1.75~T, our data suggest a phase transition involving the VL. Thus, establishing support to the previous interpretation that phase transitions of the VL occur at high fields below $H_{c_2}$, a claim based solely on results of electrical resistivity and magnetization measurements \cite{Sato, Tomy}. At low fields, our data provide a direct evidence for a disorder-induced transition into a vortex-glass at a critical field of $\mu_{0}H_{l}$~$\approx$~700~G. 

The longitudinal and transverse correlation lengths, $\xi_{L}$ and $\xi_{T}$, refer to the directions parallel and perpendicular to the external field, respectively. In our case, $\xi_{T}$ is related to the width of the Bragg spots observed directly on the 2D SANS detector, fitted using a two-dimensional Gaussian. $\xi_{L}$ is directly related to the longitudinal width, i.e., the width of the rocking curve (see below).

Single-crystalline samples were synthesized by means of a Sn self-flux method. High-purity elements (with atomic parts of three Yb, four Rh and a Sn flux) were molten at 1050~$^{\circ}$C in an evacuated and sealed quartz tube. After two hours at that temperature, the mixture was cooled to 520~$^{\circ}$C with a cooling rate of 4 ~$^{\circ}$C/h and then quenched to room temperature. The excess tin flux was removed using diluted HCl. The single crystals were characterized by means of x-ray Laue diffraction, electrical resistivity and heat capacity \cite{Danielchar}. We find that our crystals are of higher quality than previously reported materials \cite{Tomy}.

The SANS investigations presented in this study were carried out using the instruments SANS-I at the SINQ neutron source at the Paul Scherrer Institute, Villigen, Switzerland and D33 at the institute Laue-Langevin, Grenoble, France. Neutron wavelengths between 4.7~$\AA$ and 17.5~$\AA$ with a wavelength spread of 10~\% were employed. The incoming neutron beam was collimated over a distance of 18~m (SANS-I) or 12.8~m (D33) and the sample was placed inside a cryomagnet with a horizontal field parallel to the incoming neutron beam. The diffracted neutron beam was detected on a two-dimensional surface detector containing of 128 x 128 pixels \cite{Kohlbrecher}. The sample (mass $m$~=~294~mg, area $A$~=~22.3~mm$^2$ and thickness $d$~=~1.4~mm) was oriented by means of x-ray Laue diffraction and mounted on a 0.5~mm thick Al-plate of high purity. 

\begin{figure}[tbh]
\includegraphics[width=\linewidth]{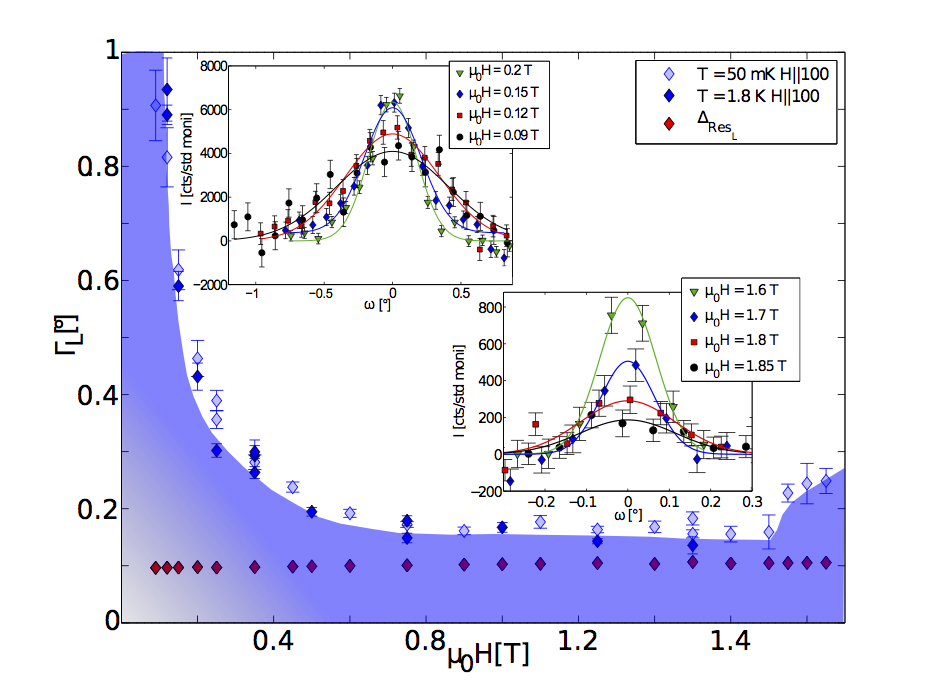}
\caption{FWHM ($\Gamma_{L}$) of the rocking curve of a representative magnetic Bragg spot at $T$~=~50~mK and 1.8~$K$ for $\vec{H}||$[100] measured on SANS-I. Red diamonds represent the longitudinal instrument resolution $\Delta_{Res_L}$. Inset: selected rocking curves at low and high magnetic fields at $T$~=~50~mK. Selected widths: $\Gamma_L$(1.6~T)~=~0.16~$\pm$~0.01~$^\circ$ and $\Gamma_L$(1.8~T)~=~0.24~$\pm$~0.03~$^\circ$.}
\label{fwhm}
\end{figure}

A rocking curve is measured by rotating the sample and the cryomagnet through a reciprocal lattice vector. This rotation angle is called rocking angle. Bragg spots appear when the Bragg condition is satisfied at the detector. In our case, background measurements were performed at $T$ = 10 K, well above $T_c$. For each magnetic field we cooled the sample from 10 K to the required temperature, while wiggling continuously the magnetic filed with an amplitude of 1 \% away from the initial value. This procedure can help to restore an ordered VL from an metastable disordered state. Typical examples of rocking curves, performed on SANS-I, are shown in the inset of Fig. \ref{fwhm}.

In our experiment the demagnetization field $\Delta B$, estimated from reported magnetization results \cite{Sato}, was neglected. For our crystal geometry $\Delta B$ is of the order of 3 \% of the applied field $\mu_{0}H$ at 1000 G and it decreases in importance for higher fields. 

A typical example of our data, measured at $\mu_{0}H$~=~3500~G and $T$~=~50~mK, is shown in Fig. \ref{phasediagramm}. The VL geometry corresponds to a slightly distorted hexagonal lattice for all investigated temperatures and field strengths \cite{Danielchar}. We note that in cubic crystals two VL domains rotated by 30$^\circ$ may be expected, due to pinning along the main axes. In our case, a single domain is found. This seems related to the lack of the fourfold rotation symmetry and shows that rather subtle details of the structure play a determining role in the VL geometry. We note that vortex lattice studies using SANS have been previously published in a symmetry related compound, PrOs$_4$Sb$_{12}$. There, it was found a substantial distortion of the vortex lattice from the ideal hexagonal lattice, which was attributed to evidence for nodes in the superconducting gaps of the order parameter \cite{Huxley2}. We find no hint for this to happen in our material.

For decreasing magnetic field strengths below approximately 1~T we observe a gradual, but dramatic, increase of the FWHM of the rocking curve $\Gamma_{L}$ (the same behavior is found for $\Gamma_{T}$). Fig. \ref{fwhm} displays the relevant data measured at SANS-I. The increase of $\Gamma_{L}$ at smaller magnetic field strengths is independent of the crystal orientation relative to the external magnetic field (data not shown). A flattening of the magnetic Bragg reflection, i.e., an increase of $\Gamma_{L}$, evidences a broadening of the reciprocal range where the Bragg condition is fulfilled and indicates a softening of the vortex lattice. Despite the increasing broadening of the magnetic Bragg spots, their position in reciprocal space follows the typical square-root behavior for an hexagonal VL down to magnetic field strengths of 350 G \cite{Danielchar}. Moreover, no anomalies are observed in the integrated intensity and the magnetic form factor at low and high magnetic fields \cite{Danielchar}. As we shall see below, our data point towards a transition of the VL into a disordered phase.

From the measured FWHM of the magnetic Bragg peaks in longitudinal ($\Gamma_{L}$) and transverse ($\Gamma_{T}$) directions we obtain the correlation lengths $\xi_{L}$ and $\xi_{T}$ \cite{alex, yaron}:

\begin{equation}
\frac{\xi_{L,T}}{a_{0}}=\frac{1}{\pi\sqrt{\Gamma^2_{L,T}-\Delta^2_{Res_{L,T}}}cos(\zeta)}.
\label{correlationlength}
\end{equation}
Here, $a_{0}$ denotes the flux line lattice spacing, $a_{0}$~=~2$\pi$/$q$, $\Gamma_{L,T}$ is the measured longitudinal or transverse width, $\Delta_{Res_{L,T}}$ describes the longitudinal or transverse instrument resolution and cos($\zeta$) is the Lorentz factor \cite{Danielchar}. The resolution of the instrument is determined by contributions of (i) the velocity selector and (ii) the collimation section. In our experiments $\Delta_{Res_{L,T}}$ are found to be of the order of 0.1$^\circ$ and 0.05$^\circ$, respectively, for both instruments, SANS-I and D33. The calculated resolution does not change appreciable for all the conditions of our experiments. The resolution correction has not been implemented for each of the data points in Fig. \ref{fwhm} in order to show the raw data. $\Delta_{Res_{L}}$ is also shown in Fig. \ref{fwhm} for the instrument configurations used on SANS-I. Therefore, $\Gamma_{L}(H)$ is not limited by the resolution of the instruments for all fields (see Fig. \ref{fwhm}). This is also the case for the larger $\Gamma_{T}(H)$. This shows that for all the applied fields the VL is not resolution limited and, in fact, it correspond to a Bragg glass as it may be expected for cases where the weak disorder cannot be neglected. 

$H_{c_2}$($T$) and the onset of the high-field transition has been suggested in Ref. \cite{Tomy}, but to our knowledge, there has been no reports on the detailed $H$-$T$ phase diagram of the superconducting state clearly indicating $H_{c_1}$ of this material. From our previous measurements, we established that $\mu_0H_{c_2}$~=~2.5~T for $T$~$\leq$~2~K \cite{Danielchar}. Recent performed magnetization measurements on Yb$_{3}$Rh$_{4}$Sn$_{13}$ reveal $\mu_0H_{c_1}$~=~135~G for $T$~$\leq$~2~K. We point out that our approach to treat neutron scattering data for a VL has been successfully used on many different materials. Recently, corrections to the Born approximation relevant to thin film have been proposed \cite{Grigoriev2010}, but they seem not relevant for our experiments.

The longitudinal and transverse correlation lengths are directly proportional to the longitudinal and radial dimensions of the correlated volume, and thus proportional to the elastic moduli $c_{44}$ and $c_{66}$ \cite{alex, yaron}:
\begin{equation}
\xi_{L}\propto c_{44}c_{66},
\label{c66}
\end{equation}
 \begin{equation}
\xi_{T}\propto \sqrt{c_{44}c_{66}^3}.
\label{c44}
\end{equation}
$c_{44}$ and $c_{66}$ are the tilt and shear modulus of the flux line lattice. Thus, the ratio $\xi_{L}/\xi_{T}$ directly reveals possible field-induced changes of $\sqrt{c_{44}/c_{66}}$.

\begin{figure}[tbh]
\includegraphics[width=\linewidth]{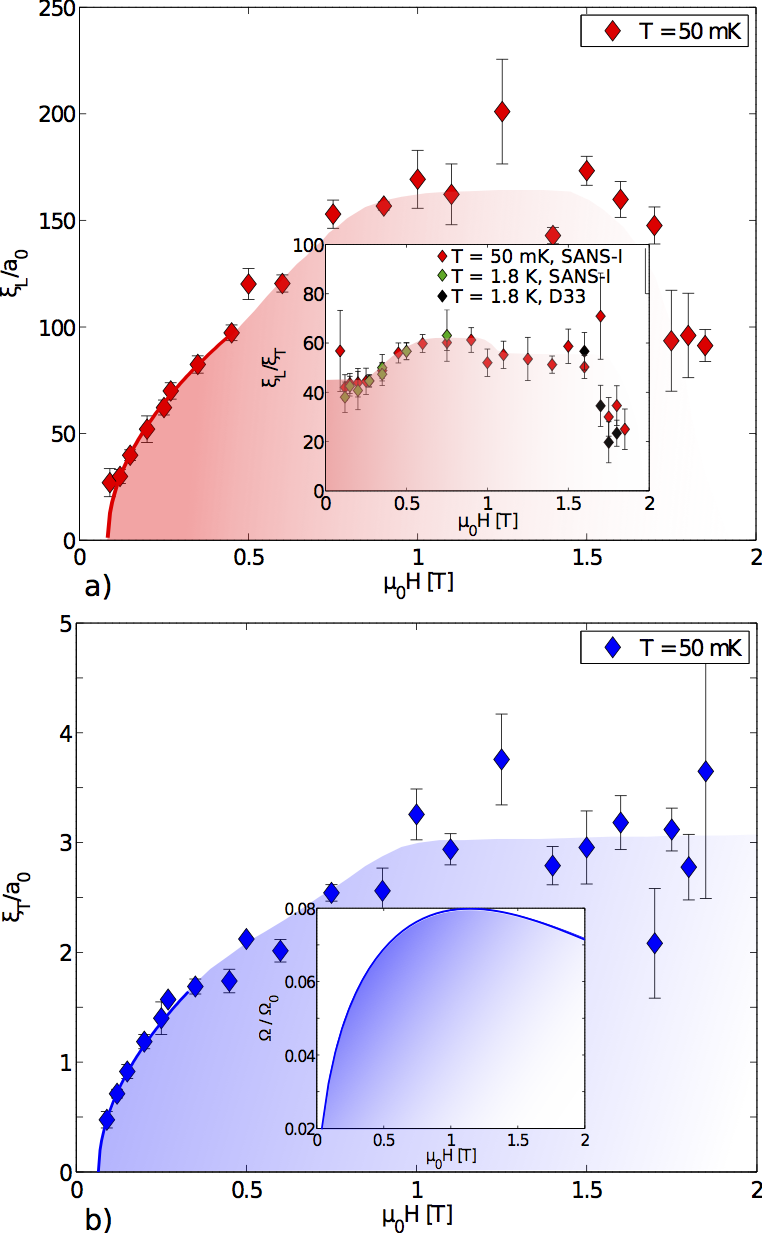}
\caption{\textbf{(a)} Normalized longitudinal correlation length $\xi_{L}/a_{0}$ as a function of the magnetic field at $T$~=~50~mK for $\vec{H}||[100]$. Inset: ratio of longitudinal and transverse correlation lengths $\xi_L/\xi_T$ $\propto$ $\sqrt{c_{44}/ c_{66}}$ measured on SANS-I and D33 at $T$~=~50~mK and 1.8~K. \textbf{(b)} Normalized transverse correlation length $\xi_{L}/a_{0}$ as a function of the field at $T$~=~50~mK for $\vec{H}||[100]$. Solid lines: $\xi_{L,T}$/$a_0$~=~$\xi_{0_{L,T}}$/$a_0\sqrt{H-H_{l}}$ with $\xi_{0_{L}}$/$a_0$~=~1.62~$\pm$ 0.05 ($\xi_{0_{T}}$/$a_0$~=~0.0318~$\pm$~0.0008) and $H_{l}$~=~700~$\pm$~75~G. Inset: simulated vortex-vortex interaction as a function of field \cite{Chaves}.}
\label{correlationlength}
\end{figure}

The normalized longitudinal and transverse correlation lengths as a function of the applied field are shown in Fig.~\ref{correlationlength}. $\xi_{L}$ is of the order of 50 times larger than $\xi_{T}$. Their relative ratio ($\sqrt{c_{44}/c_{66}}$) stays roughly constant for intermediate fields, between 1 and just below 1.7 T (see inset of Fig. \ref{correlationlength} a). In this regime, the intermediate fields, the VL is most stable, revealing the maximum normalized correlation lengths, $\xi_{L}/a_{0}$ and $\xi_{T}/a_{0}$. Here, the normalized correlation lengths change little ($\xi_{L}$ $\approx$ 161$a_{0}$ and $\xi_{T}$~$\approx$~3$a_{0}$). 

At field strengths, larger than 1.7~T, the normalized correlation length $\xi_L/a_0$ undergoes a sharp decrease to an intermediate value in the form of a jump, not observed in $\xi_T/a_0$. The ratio $\xi_{L}/\xi_{T}$ also displays an abrupt drop for fields of the order of 1.7 T (see Fig. \ref{correlationlength} a). The same behavior is observed on both instruments SANS-I and D33, with different wavelengths and varying collimation sections. This shows that the observed effect is intrinsic and does not dependent on the configuration of the instrument.  The jump evidences a sharp, but partial softening of the elastic tilt modulus $c_{44}$. The field-induced changes of $c_{66}$ are weaker, if any, and likely opposite to $c_{44}$ (see Eqs. \ref{c66} and \ref{c44}). We point out that in this field region the normalized vortex-vortex interaction displays only a modest decrease of only a few percent (see inset Fig. \ref{correlationlength}~b).  

Due to the weak SANS signal in this region, it is very difficult to follow $\xi_{L}$ and $\xi_{T}$ at higher fields. Nevertheless, the data seem reliable. They are consistent for different instruments and different configurations. Therefore, we conclude that $\xi_{L}/a_{0}$ vanishes in a complex manner as $\mu_{0}H$ approaches $\mu_{0}H_{c_2}$~$\approx$~2.5~T. Note that although $\xi_L/a_0$ shows a clear decrease in the form of a jump, near 1.7~T, this is not accompanied by a corresponding decrease of $\xi_T/a_0$, although both should vanish at $H_{c_2}$. We note that a simple melting of the Bragg glass would show a different behavior \cite{Klein}. All this is consistent with a phase transition of the vortex lattice into an intermediate glassy phase. Our data suggest the presence of further steps at higher fields that should result in a vanishing of both, $\xi_{L}/a_{0}$ and $\xi_{T}/a_{0}$, at $H_{c_2}$. But this cannot be detected in our experiments due to the rapid weakening of the SANS signal with increasing external magnetic field. The small SANS signal at high fields complicates the interpretation of the results. However, our data adds to the thermal results of Ref. \cite{Tomy, Chui, Tang, Kierfeld} and all together strongly hint for a phase transition around $\mu_{0}H$~=~1.7~T. From the same results it has been suggested that a multiple-step vortex-glass transition at fields of the order of $H_{c_{2}}$, is triggered by a proliferation of dislocations. The dislocations would finally transform the VL into a disordered glass \cite{Tomy, Chui, Tang, Kierfeld}. A  type of dislocations different from planar, for instance, of 'screw type' seems here to be less likely.

With decreasing fields below 1 T we observe a gradual reduction of $\xi_{L}/a_{0}$ and $\xi_{T}/a_{0}$, which evolves into a rapid softening of the VL at low fields. An extrapolation of the experimental $\xi_{L}/a_{0}$ and $\xi_{T}/a_{0}$ to low fields leads to vanishing correlation lengths, in a square-root manner, at about 700~G. In Fig. \ref{correlationlength} the best fits to the data using $\xi_{L,T}$/$a_0$~=~$\xi_{0_{L,T}}$/$a_0\sqrt{H-H_{l}}$ with $\xi_{0_{L}}$/$a_0$~=~1.62~$\pm$~0.05~($\xi_{0_{T}}$/$a_0$~=~0.0318~$\pm$~0.0008) and $H_{l}$~=~700~$\pm$~75 G. This value is substantially higher than our measured lower critical field $\mu_{0}H_{c_{1}}$~=~135~G (data not shown). Since our data for $\Gamma_{L}$ and $\Gamma_{T}$ (see Fig. \ref{fwhm}) show no apparent temperature effects, the inferred transition must be driven by an increasingly weaker intervortex interaction and by fixed random pinning, as we discuss below. The ratio of the two correlation lengths $\xi_{L}/\xi_{T}$ remains constant (or increases) for decreasing fields below 900~G as seen in the inset of Fig. \ref{correlationlength} a), indicating a vanishing of $c_{66}$ and $c_{44}$. We conclude that the softening of the VL reveals a typical transition of the VL into a disordered phase at low fields.

To understand the usual behavior of the correlation lengths we need to consider the vortex-vortex interaction. This interaction has been studied for many years and it is known, for instance, that it is attractive for type-I, but repulsive for type-II superconductors. For large vortex-vortex distances it may be approximated by Bessel functions \cite{Kramer}. Using modern techniques, the vortex-vortex interaction force can be calculated within the Ginzburg-Landau theory and solved numerically \cite{Chaves}. Yb$_{3}$Rh$_{4}$Sn$_{13}$ is a strong type-II superconductor with $\kappa$~=~25  \cite{Danielchar}. The vortex-vortex interaction force as a function of the external magnetic field, $H$, $\Omega(H)$ of Yb$_{3}$Rh$_{4}$Sn$_{13}$ is simulated and displayed in the inset of Fig. \ref{correlationlength}~b). $\Omega(H)$ reveals a broad maximum at around 1.2~T. For field strengths higher than 1.2~T or lower than 1~T the force gradually decreases. Since the vortex-vortex interaction only changes modestly in this region, we expect  a gradual and monotonic decrease of the correlation length with increasing field. These changes are expected to be appreciable only in the neighborhood of $H_{c_2}$($T$) due to fluctuations of the superconducting order parameter.

\begin{figure}[tbh]
\includegraphics[width=\linewidth]{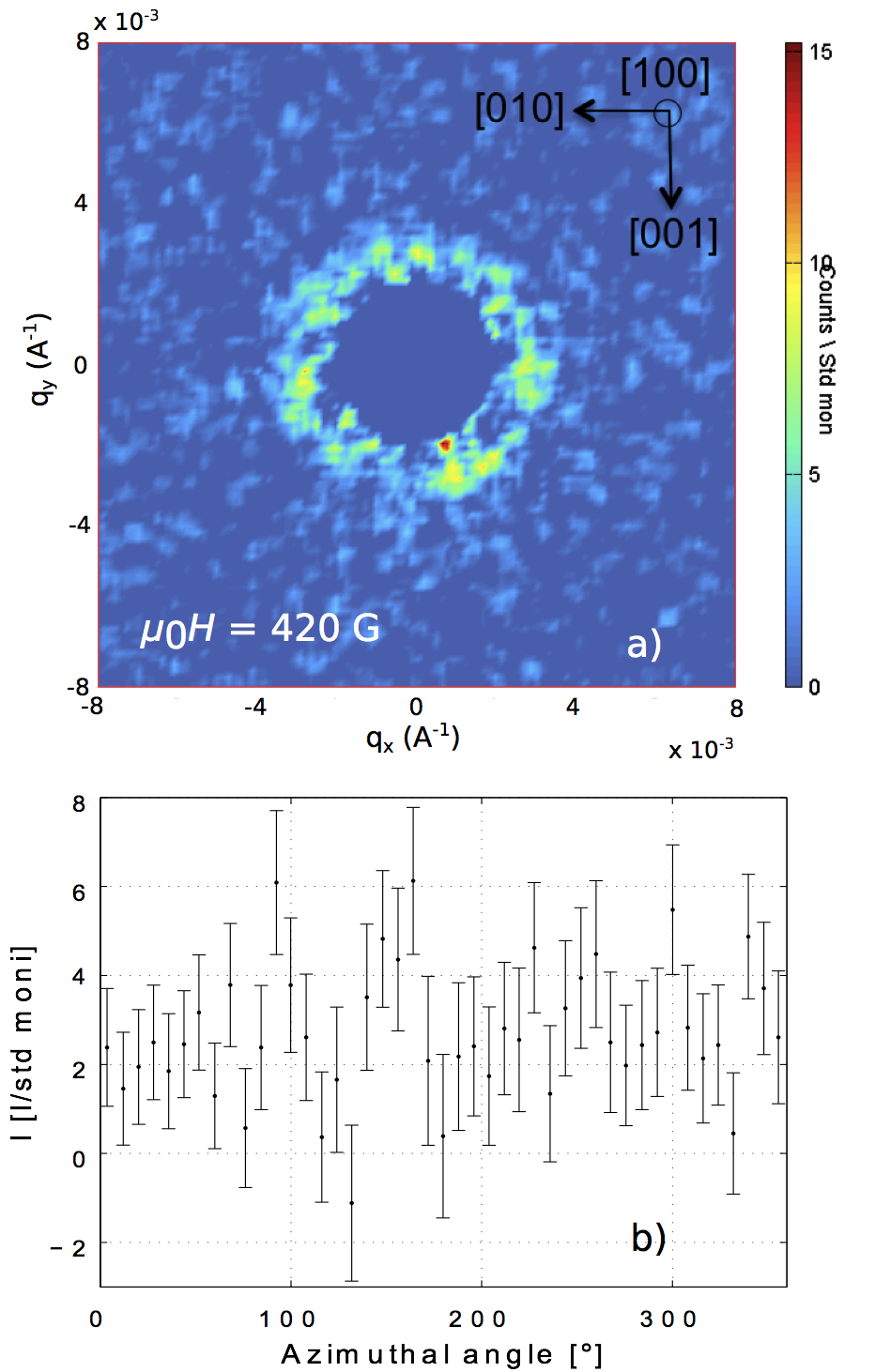}
\caption{\textbf{a)} Diffraction pattern at 420~G \textbf{b)} Azimuthal angle dependent intensity of the VL at 420~G.}
\label{azi}
\end{figure}

The transition of the VL results from a delicate balance of the vortex-vortex interaction and random forces due to pinning. The first interaction favors the VL, while the second tends to destroy it. With decreasing fields near $H_l$, the transition at low fields, the pinning forces gain the upper hand due to the dramatic decrease of $\Omega$ at low fields. This leads to a VL transition at $H_l$. A rather modest decrease of the vortex-vortex interaction is also found at high fields, but this is clearly less relevant here. Finally, we notice that by comparing data at 0.05 and 1.8~K at low fields, we conclude that thermal fluctuations may play only a minor role (if any) in the destruction of the VL. This is in accordance with the estimated very small Ginzburg number $Gi$, which provides a measure of the temperature region $\Delta$~$T/T$, near the transition temperature, for which thermal fluctuations are important. We find $Gi$~$\sim$~10$^{-8}$. Therefore, thermal fluctuations are not the driving force of the observations found at low fields.

Representative examples of the low-field SANS data are shown in Fig. \ref{azi}. Lowering the magnetic field causes a decrease in the signal-to-noise ratio in such a way, that the six Bragg spots can not be resolved for fields below $\mu_{0}H_{l}$~$\approx$~700~G (see Fig. \ref{azi} a and b). In this region the individual Bragg spots disappear and one observes only a ring in the SANS signal. This supports our claim of a transition into a glassy phase where the shear and tilt modulus disappear. The ring of SANS signal reveals the existence of short-range correlations in the disordered state. We emphasize the clear difference with the high-field occurrence. There one observes the Bragg spots at all fields and these hardly change in transverse width, as shown in Fig. \ref{correlationlength} b). Our data reveal anomalies in the longitudinal and transverse correlation lengths. They are explained by assuming transitions of the VL unto glassy states. We finally point out that in spite of the different sizes of the SANS signals at low and high fields, as the reader may appreciate in the inset of Fig. \ref{fwhm}, the claimed broadening of the Bragg reflections is intrinsic and not an artifact of the statistics.

In summary,  the VL of Yb$_3$Rh$_4$Sn$_{13}$ is well defined with a slightly distorted hexagonal geometry for the fields 700~$\leq$~$\mu_{0}H$~$<$~1.7~T. At high fields a destruction of the VL is inferred at $\mu_{0}H_h$~$\sim$~1.7 T~$<$ $\mu_{0}H_{c_2}$~$\approx$~2.5~T. Here, it is $c_{44}$, the elastic modulus for the tilt of the flux line, which cancels (in a complex manner, but more rapidly than the shear modulus $c_{66}$). This may be interpreted as evidence that dislocation planes appear with increasing number at higher fields. They lead to the destruction of the VL in a complex manner. At low fields, the magnetic Bragg spots of the VL show a gradual and temperature-independent increase of $\Gamma_{L}$ and $\Gamma_{T}$ for decreasing magnetic field strengths below 1 T. This means that the normalized correlation length $\xi_{L}/a_{0}$ and $\xi_{T}/a_{0}$ gradually decrease for decreasing fields below 1 T and extrapolates smoothly to zero at $\mu_{0}H_{l}$~$\approx$~700~G. For fields $\mu_{0}H_{c_1}$~=~135~G~$<$~$\mu_{0}H$~$<$~$\mu_{0}H_l$, the individual Bragg spots cannot be resolved but a ring of SANS signal remains. The elastic moduli $c_{66}$ and $c_{44}$ vanish at $\mu_{0}H_{l}$. All this evidences a disorder-induced (due to random pinning) vortex transition.

We would like to thank Prof. G. Blatter, Dr. V. Geshkenbein and Prof. E. Forgan for helpful discussions. We acknowledge the Swiss National Foundation for the support for two of us: D. M. with the project number 200021\_14707. The research leading to these results has received funding from the European Community's Seventh Framework Programme (FP7/2007-2013) under grant agreement number 290605  (COFUND: PSI-FELLOW). In addition, one of us (M.R.) has been supported by the Erasmus Mundus program MaMaSELF (grant 2008, Partnertship agreement 2010-0138/001) of the European Community.
\bibliographystyle{apsrev}

\bibliography{yb3rh4sn13}

\begin{thebibliography}{20}
\expandafter\ifx\csname natexlab\endcsname\relax\def\natexlab#1{#1}\fi
\expandafter\ifx\csname bibnamefont\endcsname\relax
  \def\bibnamefont#1{#1}\fi
\expandafter\ifx\csname bibfnamefont\endcsname\relax
  \def\bibfnamefont#1{#1}\fi
\expandafter\ifx\csname citenamefont\endcsname\relax
  \def\citenamefont#1{#1}\fi
\expandafter\ifx\csname url\endcsname\relax
  \def\url#1{\texttt{#1}}\fi
\expandafter\ifx\csname urlprefix\endcsname\relax\def\urlprefix{URL }\fi
\providecommand{\bibinfo}[2]{#2}
\providecommand{\eprint}[2][]{\url{#2}}

\bibitem[{\citenamefont{Blatter}(1994)\citenamefont{Blatter}}]{Blatter}
\bibinfo{author}{\bibfnamefont{G.}~\bibnamefont{Blatter}},
  \bibinfo{author}{\bibfnamefont{M.}~\bibnamefont{Feigel'man}},
  \bibinfo{author}{\bibfnamefont{V. B.}~\bibnamefont{Geshkenbein}},
  \bibinfo{author}{\bibfnamefont{A. I.}~\bibnamefont{Larkin}}, \bibnamefont{and}
  \bibinfo{author}{\bibfnamefont{V. M.}~\bibnamefont{Vinokur}},
  \bibinfo{journal}{Rev. Mod. Phys.} \textbf{\bibinfo{volume}{66}},
  \bibinfo{pages}{1125} (\bibinfo{year}{1994}).
  
  \bibitem[{\citenamefont{LeDoussal}(2010)\citenamefont{LeDoussal}}]{LeDoussal}
\bibinfo{author}{\bibfnamefont{P.}~\bibnamefont{Le Doussal}}
  \bibinfo{journal}{Int. J. Mod. Phys. B.} \textbf{\bibinfo{volume}{24}},
  \bibinfo{pages}{3855} (\bibinfo{year}{2010}).
  
    \bibitem[{\citenamefont{Brandt}(1989)\citenamefont{Brandt}}]{Brandt}
\bibinfo{author}{\bibfnamefont{E. H.}~\bibnamefont{Brandt}}
  \bibinfo{journal}{Phys. Rev. Lett.} \textbf{\bibinfo{volume}{63}},
  \bibinfo{pages}{1106} (\bibinfo{year}{1989}).

\bibitem[{\citenamefont{Lee}(2006)\citenamefont{Lee}}]{Lee}
\bibinfo{author}{\bibfnamefont{C.}~\bibnamefont{Lee}},
  \bibinfo{author}{\bibfnamefont{T. J.}~\bibnamefont{Alexander}}, \bibnamefont{and}
  \bibinfo{author}{\bibfnamefont{Y. S.}~\bibnamefont{Kivshar}},
  \bibinfo{journal}{Phys. Rev. Lett.} \textbf{\bibinfo{volume}{97}},
  \bibinfo{pages}{180408} (\bibinfo{year}{2006}).

\bibitem[{\citenamefont{remeika}(1980)\citenamefont{Remeika, Espinosa, Cooper, Barz, Rowell, McWhan, Vandenberg, Fiske, Woolf, Hamaker, Maple, Shirane
  and Thomlinson}}]{Remeika}
\bibinfo{author}{\bibfnamefont{J. P.}~\bibnamefont{Remeika}},
  \bibinfo{author}{\bibfnamefont{G. P.}~\bibnamefont{Espinosa}},
  \bibinfo{author}{\bibfnamefont{A. S.}~\bibnamefont{Cooper}},
  \bibinfo{author}{\bibfnamefont{H.}~\bibnamefont{Barz}},
  \bibinfo{author}{\bibfnamefont{J. M.}~\bibnamefont{Rowell}},
  \bibinfo{author}{\bibfnamefont{D. B.}~\bibnamefont{McWhan}},
  \bibinfo{author}{\bibfnamefont{J. M.}~\bibnamefont{Vandenberg}},
  \bibinfo{author}{\bibfnamefont{Z.}~\bibnamefont{Fiske}},
  \bibinfo{author}{\bibfnamefont{L. D.}~\bibnamefont{Woolf}},
  \bibinfo{author}{\bibfnamefont{C.}~\bibnamefont{Hamaker}},
  \bibinfo{author}{\bibfnamefont{M. B.}~\bibnamefont{Maple}},
  \bibinfo{author}{\bibfnamefont{G.}~\bibnamefont{Shirane}}, \bibnamefont{and}
  \bibinfo{author}{\bibfnamefont{W.}~\bibnamefont{Thomlinson}},
  \bibinfo{journal}{Solid State Comm.} \textbf{\bibinfo{volume}{34}},
  \bibinfo{pages}{923} (\bibinfo{year}{1980}).

  
  \bibitem[{\citenamefont{klintberg}(2012)\citenamefont{Klintberg}}]{Klintberg}
\bibinfo{author}{\bibfnamefont{L. E.}~\bibnamefont{Klintberg}},
  \bibinfo{author}{\bibfnamefont{S. K.}~\bibnamefont{Goh}},
  \bibinfo{author}{\bibfnamefont{P. L.}~\bibnamefont{Alireza}},
  \bibinfo{author}{\bibfnamefont{P. J.}~\bibnamefont{Saines}},
  \bibinfo{author}{\bibfnamefont{D. A.}~\bibnamefont{Tompsett}},
  \bibinfo{author}{\bibfnamefont{P. W.}~\bibnamefont{Logg}},
  \bibinfo{author}{\bibfnamefont{J.}~\bibnamefont{Yang}},
  \bibinfo{author}{\bibfnamefont{B.}~\bibnamefont{Chen}},
  \bibinfo{author}{\bibfnamefont{K.}~\bibnamefont{Yoshimura}}, \bibnamefont{and}
  \bibinfo{author}{\bibfnamefont{F. M.}~\bibnamefont{Grosche}},
  \bibinfo{journal}{Phys. Rev. Lett.} \textbf{\bibinfo{volume}{109}},
  \bibinfo{pages}{237008} (\bibinfo{year}{2012}).

\bibitem[{\citenamefont{gerber}(2013)\citenamefont{Gerber}}]{Gerber}
\bibinfo{author}{\bibfnamefont{S.}~\bibnamefont{Gerber}},
  \bibinfo{author}{\bibfnamefont{J. L.}~\bibnamefont{Gavilano}},
  \bibinfo{author}{\bibfnamefont{M.}~\bibnamefont{Medarde}},
  \bibinfo{author}{\bibfnamefont{V.}~\bibnamefont{Pomjakushin}},
  \bibinfo{author}{\bibfnamefont{C.}~\bibnamefont{Baines}},
  \bibinfo{author}{\bibfnamefont{E.}~\bibnamefont{Pomjakushina}},
  \bibinfo{author}{\bibfnamefont{K.}~\bibnamefont{Conder}}, \bibnamefont{and}
  \bibinfo{author}{\bibfnamefont{M.}~\bibnamefont{Kenzelmann}},
  \bibinfo{journal}{Phys. Rev. B.} \textbf{\bibinfo{volume}{88}},
  \bibinfo{pages}{104505} (\bibinfo{year}{2013}).

\bibitem[{\citenamefont{levett}(2003)\citenamefont{Levett}}]{Levett}
\bibinfo{author}{\bibfnamefont{S. J.}~\bibnamefont{Levett}}, PhD thesis, University of Warwick, 2003.

\bibitem[{\citenamefont{miraglia}(1986)\citenamefont{miraglia}}]{miraglia}
\bibinfo{author}{\bibfnamefont{S}~\bibnamefont{Miraglia}},
  \bibinfo{author}{\bibfnamefont{J. L.}~\bibnamefont{Hodeau}}, 
  \bibinfo{author}{\bibfnamefont{M.}~\bibnamefont{Marezio}}, 
  \bibinfo{author}{\bibfnamefont{C.}~\bibnamefont{Laviron}},
  \bibinfo{author}{\bibfnamefont{M.}~\bibnamefont{Ghedira}},  \bibnamefont{and}
  \bibinfo{author}{\bibfnamefont{G. P.}~\bibnamefont{Espinosa}},
  \bibinfo{journal}{J. Solid State Chem.}\textbf{\bibinfo{volume}{63}},
  \bibinfo{pages}{358} (\bibinfo{year}{1986}).

\bibitem[{\citenamefont{Wang}(2012)\citenamefont{Wang}}]{Wang}
\bibinfo{author}{\bibfnamefont{K}~\bibnamefont{Wang}}, \bibnamefont{and}
  \bibinfo{author}{\bibfnamefont{C.}~\bibnamefont{Petrovic}},
  \bibinfo{journal}{Phys. Rev. B} \textbf{\bibinfo{volume}{86}},
  \bibinfo{pages}{024522} (\bibinfo{year}{2012}).

\bibitem[{\citenamefont{Sato}(1995)\citenamefont{Sato}}]{Sato}
\bibinfo{author}{\bibfnamefont{H.}~\bibnamefont{Sato}},
  \bibinfo{author}{\bibfnamefont{Y.}~\bibnamefont{Aoki}}, \bibnamefont{and}
  \bibinfo{author}{\bibfnamefont{H.}~\bibnamefont{Sugawara}},
  \bibinfo{journal}{J. Phys. Soc. Jpn.} \textbf{\bibinfo{volume}{64}},
  \bibinfo{pages}{3175} (\bibinfo{year}{1995}).

\bibitem[{\citenamefont{Tomy}(1997)\citenamefont{Tomy}}]{Tomy}
\bibinfo{author}{\bibfnamefont{C. V.}~\bibnamefont{Tomy}},
  \bibinfo{author}{\bibfnamefont{G.}~\bibnamefont{Balakrishnan}}, \bibnamefont{and}
  \bibinfo{author}{\bibfnamefont{D. McK.}~\bibnamefont{Paul}},
  \bibinfo{journal}{Physica C} \textbf{\bibinfo{volume}{280}},
  \bibinfo{pages}{1} (\bibinfo{year}{1997}).
 
  \bibitem[{\citenamefont{Danielchar}(2014)\citenamefont{Danielchar}}]{Danielchar}
\bibinfo{author}{\bibfnamefont{D.}~\bibnamefont{Mazzone}},
  \bibinfo{author}{\bibfnamefont{J. L.}~\bibnamefont{Gavilano}},
  \bibinfo{author}{\bibfnamefont{R.}~\bibnamefont{Sibille}},
  \bibinfo{author}{\bibfnamefont{M.}~\bibnamefont{Ramakrishnan}}, \bibnamefont{and}
  \bibinfo{author}{\bibfnamefont{M.}~\bibnamefont{Kenzelmann}},
 \bibinfo{journal}{Phys. Rev. B} \textbf{\bibinfo{volume}{90}},
  \bibinfo{pages}{020507(R)} (\bibinfo{year}{2014}).

\bibitem[{\citenamefont{Kohlbrecher}(200)\citenamefont{Kohlbrecher}}]{Kohlbrecher}
\bibinfo{author}{\bibfnamefont{J.}~\bibnamefont{Kohlbrecher}}, \bibnamefont{and}
  \bibinfo{author}{\bibfnamefont{W.}~\bibnamefont{Wagner}},
  \bibinfo{journal}{Appl. Cryst.} \textbf{\bibinfo{volume}{33}},
  \bibinfo{pages}{804} (\bibinfo{year}{2000}).
  
  \bibitem[{\citenamefont{Huxley2}(2004)\citenamefont{Huxley2}}]{Huxley2}
\bibinfo{author}{\bibfnamefont{A. D.}~\bibnamefont{Huxley}},
  \bibinfo{author}{\bibfnamefont{M. A.}~\bibnamefont{Measson}},
  \bibinfo{author}{\bibfnamefont{K.}~\bibnamefont{Izawa}},
  \bibinfo{author}{\bibfnamefont{C. D.}~\bibnamefont{Dewhurst}},
  \bibinfo{author}{\bibfnamefont{R.}~\bibnamefont{Cubitt}},
  \bibinfo{author}{\bibfnamefont{B.}~\bibnamefont{Grenier}},
  \bibinfo{author}{\bibfnamefont{H.}~\bibnamefont{Sugawara}},
  \bibinfo{author}{\bibfnamefont{J.}~\bibnamefont{Flouquet}},
  \bibinfo{author}{\bibfnamefont{Y.}~\bibnamefont{Matsuda}}, \bibnamefont{and}
  \bibinfo{author}{\bibfnamefont{H.}~\bibnamefont{Sato}}
  \bibinfo{journal}{Phys. Rev. Lett.} \textbf{\bibinfo{volume}{93}},
  \bibinfo{pages}{187005} (\bibinfo{year}{2004}).
  
  
  \bibitem[{\citenamefont{Klein}(2001)\citenamefont{Klein}}]{Klein}
\bibinfo{author}{\bibfnamefont{T.}~\bibnamefont{Klein}},
  \bibinfo{author}{\bibfnamefont{I.}~\bibnamefont{Joumard}},
  \bibinfo{author}{\bibfnamefont{S.}~\bibnamefont{Blanchard}},
  \bibinfo{author}{\bibfnamefont{J.}~\bibnamefont{Marcus}},
  \bibinfo{author}{\bibfnamefont{R.}~\bibnamefont{Cubitt}},
  \bibinfo{author}{\bibfnamefont{T.}~\bibnamefont{Glamarchi}}, \bibnamefont{and}
  \bibinfo{author}{\bibfnamefont{P.}~\bibnamefont{Le Doussai}},
  \bibinfo{journal}{Letters to Nature} \textbf{\bibinfo{volume}{413}},
  \bibinfo{pages}{404} (\bibinfo{year}{2001}).
  
  \bibitem[{\citenamefont{Chui}(1992)\citenamefont{Chui}}]{Chui}
\bibinfo{author}{\bibfnamefont{S. T.}~\bibnamefont{Chui}},
  \bibinfo{journal}{Europhys. Lett} \textbf{\bibinfo{volume}{20}},
  \bibinfo{pages}{535} (\bibinfo{year}{2001}).
  
    \bibitem[{\citenamefont{Tang}(1996)\citenamefont{Tang}}]{Tang}
\bibinfo{author}{\bibfnamefont{C.}~\bibnamefont{Tang}},
\bibinfo{author}{\bibfnamefont{X.}~\bibnamefont{Ling}},
\bibinfo{author}{\bibfnamefont{S.}~\bibnamefont{Bhattacharya}}, \bibnamefont{and}
  \bibinfo{author}{\bibfnamefont{P.}~\bibnamefont{Chaikin}},
  \bibinfo{journal}{Europhys. Lett} \textbf{\bibinfo{volume}{35}},
  \bibinfo{pages}{597} (\bibinfo{year}{1996}).
  
   \bibitem[{\citenamefont{Kierfeld}(2000)\citenamefont{Kierfeld}}]{Kierfeld}
\bibinfo{author}{\bibfnamefont{J.}~\bibnamefont{Kierfeld}}, \bibnamefont{and}
  \bibinfo{author}{\bibfnamefont{V.}~\bibnamefont{Vinokur}},
  \bibinfo{journal}{Phys. Rev. B} \textbf{\bibinfo{volume}{61}},
  \bibinfo{pages}{14928} (\bibinfo{year}{2000}).
  
    \bibitem[{\citenamefont{alex}(2014)\citenamefont{alex}}]{alex}
\bibinfo{author}{\bibfnamefont{A. T.}~\bibnamefont{Holmes}},
   \bibinfo{journal}{Phys. Rev. B} \textbf{\bibinfo{volume}{90}},
  \bibinfo{pages}{024514} (\bibinfo{year}{2014}).

  \bibitem[{\citenamefont{yaron}(1995)\citenamefont{yaron}}]{yaron}
\bibinfo{author}{\bibfnamefont{U.}~\bibnamefont{Yaron}},
  \bibinfo{author}{\bibfnamefont{P. L.}~\bibnamefont{Gammel}},
  \bibinfo{author}{\bibfnamefont{D. A.}~\bibnamefont{Huse}},
  \bibinfo{author}{\bibfnamefont{J.}~\bibnamefont{Marcus}},
  \bibinfo{author}{\bibfnamefont{R. N.}~\bibnamefont{Kleiman}},
  \bibinfo{author}{\bibfnamefont{C. S.}~\bibnamefont{Oglesby}},
  \bibinfo{author}{\bibfnamefont{E.}~\bibnamefont{Bucher}},
    \bibinfo{author}{\bibfnamefont{B.}~\bibnamefont{Batlogg}},
      \bibinfo{author}{\bibfnamefont{D. J.}~\bibnamefont{Bishop}},
  \bibinfo{author}{\bibfnamefont{K.}~\bibnamefont{Mortensen}}, \bibnamefont{and}
  \bibinfo{author}{\bibfnamefont{K. N.}~\bibnamefont{Clausen}},
  \bibinfo{journal}{Nature} \textbf{\bibinfo{volume}{376}},
  \bibinfo{pages}{753} (\bibinfo{year}{1995}).

  \bibitem[{\citenamefont{Grigoriev2010}(2010)\citenamefont{Grigoriev2010}}]{Grigoriev2010}
\bibinfo{author}{\bibfnamefont{A.}~\bibnamefont{Chaves}},
  \bibinfo{author}{\bibfnamefont{S. V.}~\bibnamefont{Grigoriev}},
  \bibinfo{author}{\bibfnamefont{A. V..}~\bibnamefont{ Syromyatnikov}},
  \bibinfo{author}{\bibfnamefont{A. P.}~\bibnamefont{Chumakov}},
  \bibinfo{author}{\bibfnamefont{N. A.}~\bibnamefont{Grigoriev}},
  \bibinfo{author}{\bibfnamefont{K. S.}~\bibnamefont{Napolskii}},
  \bibinfo{author}{\bibfnamefont{I. V.}~\bibnamefont{Roslyakov}},
  \bibinfo{author}{\bibfnamefont{A. A.}~\bibnamefont{Eliseev}},
  \bibinfo{author}{\bibfnamefont{A. V.}~\bibnamefont{Petukhov}}, \bibnamefont{and}
  \bibinfo{author}{\bibfnamefont{H.}~\bibnamefont{Eckerlebe}}
  \bibinfo{journal}{Phys. Rev. B} \textbf{\bibinfo{volume}{81}},
  \bibinfo{pages}{125405} (\bibinfo{year}{2010}).

  \bibitem[{\citenamefont{Kramer}(1971)\citenamefont{Kramer}}]{Kramer}
\bibinfo{author}{\bibfnamefont{L.}~\bibnamefont{Kramer}}
  \bibinfo{journal}{Phys. Rev. B} \textbf{\bibinfo{volume}{3}},
  \bibinfo{pages}{3821} (\bibinfo{year}1971).

  \bibitem[{\citenamefont{Chaves}(2011)\citenamefont{Chaves}}]{Chaves}
\bibinfo{author}{\bibfnamefont{A.}~\bibnamefont{Chaves}},
  \bibinfo{author}{\bibfnamefont{F. M.}~\bibnamefont{Peeters}},
  \bibinfo{author}{\bibfnamefont{G. A.}~\bibnamefont{Farias}}, \bibnamefont{and}
  \bibinfo{author}{\bibfnamefont{M. V.}~\bibnamefont{Milo\u{s}evi\'{c}}}
  \bibinfo{journal}{Phys. Rev. B} \textbf{\bibinfo{volume}{83}},
  \bibinfo{pages}{054516} (\bibinfo{year}{2011}).


\end{thebibliography}

\end{document}